\documentstyle[amssymb,aps,preprint]{revtex}

\begin{document}
\title{The effect of pressure on statics, dynamics and stability of multielectron
bubbles}
\author{J. Tempere$^{1,2}$, Isaac F. Silvera$^{1}$, J. T. Devreese$^{2}$}
\address{$^{1}$ Dept. Physics, Harvard University, 14-A Oxford Street, Cambridge MA
02139, USA.\\
$^{2}$ Dept. Natuurkunde, Universiteit Antwerpen, Universiteitsplein 1,
B2610 Antwerpen, Belgium.}
\date{August 23, 2001}
\maketitle

\begin{abstract}
The effect of pressure and negative pressure on the modes of oscillation of
a multi-electron bubble in liquid helium is calculated. Already at low
pressures of the order of 10-100 mbar, these effects are found to
significantly modify the frequencies of oscillation of the bubble.
Stabilization of the bubble is shown to occur in the presence of a small
negative pressure, which expands the bubble radius. Above a threshold
negative pressure, the bubble is unstable.
\end{abstract}


\bigskip

\smallskip

Multielectron bubbles (MEBs) in liquid helium are fascinating entities
expected to display\ novel resonant behavior, the possibility of
superconductivity, and sufficient electron density for Wigner
crystallization and quantum melting. MEBs are bubbles inside the helium,
containing only electrons that\ form a curved two-dimensional electron gas
(2DEG) on the spherical surface of the bubble, the width of the thin
spherical shell that conforms to the helium surface being of order 5-20 \AA\ 
\cite{ShikinJETP27,SalomaaPRL47}, compared to 70-80 \AA\ on a flat surface
\cite{ColeRMP46}. New efforts to trap and localize MEBs for long periods of
time \cite{SilveraBAPS46} have led to further consideration of the long term
stability of bubbles. In this letter we discuss the use of pressure, and in
particular negative pressure to stabilize MEBs against dynamic instabilities.

The radius of the bubble depends on the enclosed charge. In a simplified
model, valid for large bubbles, the radius is given by $%
R_{C}=[e^{2}N^{2}/(16\pi \sigma \varepsilon )]^{1/3}$\ where $e$\ is the
electron charge, $N$ is the number of electrons in the bubble, $\sigma $\ is
the surface tension of helium and $\varepsilon $\ is the dielectric constant
of helium \cite{ShikinJETP27}. A single electron bubble has a radius of 17.2 
\AA\ \cite{GrimesPRB41} while, for example, a bubble of 10$^{4}$ electrons
has a theoretical radius of 1 micron. There is some question about the
static stability of an MEB: since the energy of the bubble, defined as the
sum of the electrostatic and the surface tension energy, is proportional to $%
N^{4/3}$ \cite{ShikinJETP27},\ clearly the energy of two bubbles with $N/2$\
electrons is lower than that of a single bubble with $N$\ electrons.
Evidently fissioning is hindered by a formation barrier, since MEBs have
been observed \cite{VolodinJETPLet26}. Gravitational fields may flatten very
large bubbles and lead to instability \cite{ShikinJETP27}. Dexter and Fowler 
\cite{DexterPR183} showed that two-electron bubbles are unstable. Salomaa
and Williams\cite{SalomaaPRL47} considered the dynamic stability against
fissioning off of single electrons from large bubbles and found stability
against this decay mode for bubbles with $N$ greater than 15-20. It is
straightforward to show that a positive pressure radially stabilizes a
bubble,\ although angular modes can be unstable.\ With increasing negative
pressure the bubble is first absolutely stabilized and then explodes, as we
shall describe. Salomaa and Williams also\ considered the dynamic
instability due to one of the surface oscillation modes, or ripplons, being
soft (zero frequency) and found that this mode may be stabilized by the
anharmonicity in the\ bubble's radial oscillation that results in a radius
larger than $R_{\text{C}}.$ In this communication we study pressure related
effects on the frequency of the modes of oscillation, on the equilibrium
bubble radius, and on the stability of MEBs; we show the counterintuitive
result that positive pressures can destabilize all higher angular modes,
while negative pressures have a window of stabilization. We neglect gravity
so that the MEBs are spherical.

The frequencies of the modes of oscillation of a charged droplet were first
calculated by Rayleigh \cite{RayleighPRS29}, and in the case of a charged
bubble by Plesset and Prosperetti \cite{PlessetARFM9}. We first set up a
Lagrangian formalism to calculate the spherical ripplon modes. We then
consider the effect of pressure on the static and dynamic properties of the
bubble. The surface of the bubble is described by a function $R(\theta
,\varphi )$ that gives the distance of the surface from the geometrical
center of the bubble, in the direction given by the two spherical angles $%
\theta ,\varphi $ . This function can be written as $R(\theta ,\varphi )=R_{%
\text{b}}+u(\theta ,\varphi ),$where $R_{\text{b}}$ is the angle-averaged
radius of the bubble, and $u(\theta ,\varphi )$ describes the deformation of
the surface from a sphere. This deformation can be expanded in a series of
spherical harmonic deformations $Y_{\ell m}(\theta ,\varphi )$ with
amplitude $Q_{\ell m}$%
\begin{equation}
u(\theta ,\varphi )=%
\mathop{\displaystyle\sum}%
\limits_{\ell =1}^{\infty }%
\mathop{\displaystyle\sum}%
\limits_{m=-\ell }^{\ell }Q_{\ell m}Y_{\ell m}(\theta ,\varphi ).
\end{equation}
In what follows, we shall assume that the amplitude of deformation is small,
so that for all $\{\ell ,m\}$, $\sqrt{\ell (\ell +1)}Q_{\ell m}/R_{\text{b}%
}\ll 1.$

The kinetic energy $T$ associated with the motion of the liquid helium
surface can be derived from the velocity potential $\psi $. The kinetic
energy for incompressible flow of an inviscid fluid is given by 
\begin{equation}
T=\frac{\rho }{2}%
\displaystyle\int %
\limits_{0}^{2\pi }d\varphi 
\displaystyle\int %
\limits_{0}^{\pi }d\theta \text{ }\psi (R_{\text{b}},\theta ,\varphi )\left[ 
{\bf n}\cdot \left. \nabla \psi (r,\theta ,\varphi )\right| _{r=R_{\text{b}}}%
\right] \text{ }R_{\text{b}}^{2}\sin \theta ,
\end{equation}
with the\ density of helium $\rho =145$ kg/m$^{3}$, and {\bf n}\ is the unit
vector normal to the bubble surface. The coefficients of expansion in
spherical harmonics of the velocity potential are then expressed as a
function of the deformation amplitude $Q_{\ell m}$, leading to 
\begin{equation}
T=%
{\displaystyle{\rho R_{\text{b}}^{3} \over 2}}%
\mathop{\displaystyle\sum}%
\limits_{\ell =0}^{\infty }%
\mathop{\displaystyle\sum}%
\limits_{m=-\ell }^{\ell }%
{\displaystyle{1 \over \ell +1}}%
\left| \dot{Q}_{\ell m}\right| ^{2}.  \label{kinetic}
\end{equation}

The potential energy of the deformed bubble results from the surface
tension, the pressure exerted on the bubble, and the electrostatic forces
between the electrons of the spherical 2DEG. The surface tension energy can
be written as $\sigma $S where $\sigma =3.6\times $10$^{-4}$ J/m$^{2}$ is
the surface tension of helium at zero pressure, and S is the area of the
deformed surface \cite{RayleighPRS29}: 
\begin{equation}
\text{S}=4\pi R_{\text{b}}^{2}+%
{\displaystyle{1 \over 2}}%
\mathop{\displaystyle\sum}%
\limits_{\ell =1}^{\infty }%
\mathop{\displaystyle\sum}%
\limits_{m=-\ell }^{\ell }(\ell ^{2}+\ell +2)\left| Q_{\ell m}\right| ^{2}.
\label{surface}
\end{equation}
Deforming the bubble will change its volume and perform $p$V work against
the external pressure $p$ of the helium liquid. The volume of the deformed
bubble up to second order in the deformation amplitudes is given by \cite
{RayleighPRS29}: 
\begin{equation}
\text{V}=%
{\displaystyle{4\pi  \over 3}}%
R_{\text{b}}^{3}+R_{\text{b}}%
\mathop{\displaystyle\sum}%
\limits_{\ell =1}^{\infty }%
\mathop{\displaystyle\sum}%
\limits_{m=-\ell }^{\ell }\left| Q_{\ell m}\right| ^{2}.  \label{volume}
\end{equation}
The electrostatic energy of the electrons in the spherical 2DEG in the MEB
can be derived by taking into consideration that the electrons are strongly
confined in the direction perpendicular to the helium surface (the binding
energy is of the order of 10 K) and anchored to that surface, but free to
move in the directions parallel to the helium surface. This leads to the
following expression for the Coulomb part $U_{\text{C}}$ of the potential
energy \cite{RayleighPRS29}: 
\begin{equation}
U_{\text{C}}=%
{\displaystyle{N^{2}e^{2} \over 2\varepsilon R_{\text{b}}}}%
-%
{\displaystyle{N^{2}e^{2} \over 8\pi \varepsilon R_{\text{b}}^{3}}}%
\mathop{\displaystyle\sum}%
\limits_{\ell =1}^{\infty }%
\mathop{\displaystyle\sum}%
\limits_{m=-\ell }^{\ell }\ell |Q_{\ell m}|^{2},  \label{coulomb}
\end{equation}
with $\varepsilon =1.0572$ the dielectric constant of helium. This
expression is valid up to second order in the deformation amplitude $Q_{\ell
m}$ and does not include exchange or correlation energies of the electron
gas.

Collecting the previous expressions (\ref{kinetic},\ref{surface},\ref{volume}%
,\ref{coulomb}) for the different energy contributions leads to the
following expression for the Lagrangian ${\cal L}_{\text{bubble}}=T-\sigma $S%
$-p$V$-U_{C}$ : 
\begin{eqnarray}
{\cal L}_{\text{bubble}} &=&%
{\displaystyle{\rho R_{\text{b}}^{3}\dot{R}_{\text{b}}^{2} \over 2}}%
-4\pi \sigma R_{\text{b}}^{2}-\frac{4\pi }{3}pR_{\text{b}}^{3}-%
{\displaystyle{N^{2}e^{2} \over 2\varepsilon R_{\text{b}}}}%
\nonumber \\
&&+%
\mathop{\displaystyle\sum}%
\limits_{\ell =1}^{\infty }%
\mathop{\displaystyle\sum}%
\limits_{m=-\ell }^{\ell }\left\{ 
{\displaystyle{\rho R_{\text{b}}^{3} \over 2}}%
{\displaystyle{1 \over \ell +1}}%
\left| \dot{Q}_{\ell m}\right| ^{2}-\left[ \frac{\sigma }{2}(\ell ^{2}+\ell
+2)+pR_{b}-%
{\displaystyle{N^{2}e^{2} \over 8\pi \varepsilon R_{\text{b}}^{3}}}%
\ell \right] |Q_{\ell m}|^{2}\right\} .  \label{Lagrangian}
\end{eqnarray}
The novel part in this Lagrangian as compared to previous treatments \cite
{SalomaaPRL47,RayleighPRS29,PlessetARFM9} lies in the terms related to the
pressure. We shall return to the harmonic solutions of this Lagrangian after
discussing some static properties.\ 

The equilibrium radius of large bubbles, in the absence of deformations, $%
Q_{\ell m}=0$, can be found by minimizing the potential energy $U=\sigma $S$%
+p$V$+U_{C}$ as a function of the bubble radius. The potential energy is: 
\begin{equation}
U(R_{\text{b}})=\frac{4\pi }{3}pR_{\text{b}}^{3}+4\pi \sigma R_{\text{b}%
}^{2}+\frac{e^{2}N^{2}}{2\varepsilon (R_{\text{b}}-d)}+\frac{N\hbar ^{2}}{%
2m_{\text{e}}d^{2}}-0.3176\frac{e^{2}N^{4/3}}{\varepsilon \left( R_{\text{b}%
}^{2}d\right) ^{1/3}}  \label{U}
\end{equation}
For large bubbles, with $N\gtrapprox 10^{3},$ only the first three terms in
the right-hand side of (\ref{U}) play a role -- these terms are also present
in the Lagrangian (\ref{Lagrangian}). The fourth term ($m_{\text{e}}$ is the
electron mass) is due to Shikin \cite{ShikinJETP27} who makes the argument
that there is a finite thickness $d\ll R$ of the electron layer that should
be taken into account, along with $R_{\text{b}},$ as a variational parameter
to minimize the potential energy$.$ The density functional calculations of
Shung and Lin \cite{ShungPRB45} show that the exchange energy of the
electron gas also plays a relevant role for small bubbles (N%
\mbox{$<$}%
1000), and that for practical purposes it is well approximated by adding an
exchange term, the fifth term in the right-hand side of (\ref{U}).

The equilibrium radius $R_{\text{eq}}$ of the MEB\ is found as a function of
the number of electrons and the exerted pressure, by minimizing the
potential energy $U(R_{\text{b}})$ with respect to $d$ and $R_{\text{b}}$.
The value of $R_{\text{b}}$\ which minimizes the potential energy $U(R_{%
\text{b}})$\ is the equilibrium radius $R_{\text{eq}}$ shown in Fig. 1 as a
function of pressure for different numbers of electrons. The pressure
reduces\ the equilibrium radius by a significant factor as compared to the
zero pressure radius. Most of the change in $R_{\text{eq}}$ occurs at low
pressures. The behavior of $R_{\text{eq}}$ as a function of pressure shown
in Fig. 1 is similar for numbers of electrons up to $10^{8}$ and\ larger.\
The graph extends to negative pressure \cite{negpress} increasing $R_{\text{%
eq}}$, since superfluid liquid helium can sustain a substantial negative
pressure. For any number of electrons in the bubble, there exists a critical
negative pressure making the bubble unstable against runaway expansion. We
found that the largest equilibrium radius that can be achieved before the
critical underpressure is reached is approximately $1.5$ times the
equilibrium bubble radius at zero pressure, for any number of electrons from
a few hundred to 10$^{8}$. When the critical underpressure is reached, the
only equilibrium radius is $R_{\text{eq}}\rightarrow \infty $. The critical
underpressure appears to be inversely proportional to the number of
electrons in the bubble, but this could not be shown explicitly.

At fixed $R_{\text{b}}$ (the equilibrium radius), the part of the Lagrangian
(\ref{Lagrangian}) pertaining to the spherical ripplon modes represents a
collection of harmonic oscillators in the coordinates $Q_{\ell m}$ with
ripplon frequencies

\begin{equation}
\omega _{\ell }=\sqrt{\frac{\ell +1}{\rho R_{\text{b}}^{3}}\left[ \sigma
(\ell ^{2}+\ell +2)+2pR_{\text{b}}-\frac{N^{2}e^{2}}{4\pi \varepsilon R_{%
\text{b}}^{3}}\ell \right] }.  \label{w(L)}
\end{equation}
These frequencies are independent of the azimuthal index $m$. Fig. 2 shows
the pressure dependence of the frequency of the spherical ripplon modes $%
\ell =1,...,10$. Note first that the $\ell =1,2$ modes have vanishing
frequencies at pressures $p\geqslant 0$. This was discussed for zero
pressure by Salomaa and Williams \cite{SalomaaPRL47}, who concluded that the 
$\ell =2$ mode can be dynamically stabilized if the effective radius of the
bubble satisfies $R_{\text{b}}>R_{\text{eq}}$ (the $\ell =1$\ corresponds to
uniform translation). However, the present treatment shows that this
argument is no longer valid at increasing pressure. In fact, subsequent
modes become unstable, as can be seen from Fig. 2 and expression (\ref{w(L)}%
). If the exchange and confinement energy terms in (\ref{U}) are neglected
(a reasonable assumption for bubbles with $N>10^{3}$), the equilibrium
radius satisfies $2pR_{\text{eq}}+4\sigma =e^{2}N^{2}/4\pi \varepsilon R_{%
\text{eq}}^{3},$ so that in this case 
\begin{equation}
R_{\text{b}}=R_{\text{eq}}\Rightarrow \omega _{\ell }=\sqrt{\frac{\ell ^{2}-1%
}{\rho R_{\text{eq}}^{3}}\left[ \sigma (\ell -2)-2pR_{\text{eq}}\right] }.
\end{equation}
The pressure at which a mode $\ell >2$ becomes unstable is $p=\sigma (\ell
-2)/(2R_{\text{eq}})$. For $N=10^{4}$, $\sigma /R_{\text{eq}}$ is of the
order of 100 mbar. Larger bubbles have even smaller critical pressures. The
vanishing of the frequencies of these modes indicates that an instability
occurs and that the quadratic approximation for the deformation is no longer
valid. Note however, that a small negative pressure tends to stabilize the $%
\ell =2$ mode - this is an alternative to Salomaa and Williams' proposal for
the stability of the $\ell =2$ mode, based on the assumption that the
effective radius $R_{\text{b}}>R_{\text{eq}}.$

In this letter, we have shown that small negative pressures can stabilize a
bubble against dynamic instability, while positive pressures can drive all
ripplon modes unstable. Since both negative and positive pressures are
easily achievable experimentally, it will be interesting if bubbles can be
created in a ''stable'' configuration that can be visually observed to study
these predictions. In the above considerations we have considered the
electrons to be a 2DEG, ignoring Wigner crystallization.

We thank S. N. Klimin, V. M. Fomin and J. Huang for useful comments and
discussions. This research has been supported by the Department of Energy,
grant DE-FG002-85ER45190, and by the\ GOA BOF UA 2000, IUAP, the FWO-V
projects Nos. G.0071.98, G.0306.00, G.0274.01, WOG WO.025.99N (Belgium), and
the ESF Programme VORTEX. J. Tempere is supported financially by the
FWO-Flanders as a `Postdoctoral Fellow of the Fund for Scientific Research -
Flanders'.

\bigskip

\section*{Figure captions}

FIG. 1. The equilibrium radius $R$ of a multi-electron bubble is shown as a
function of pressure for bubbles for several values of $N$. In the inset,
the thickness $d$ of the spherical electron shell in the bubble is shown as
a function of pressure for the same numbers of electrons. Both the
equilibrium radius and the thickness were obtained by minimizing the
potential energy (\ref{U}). Negative pressure can be applied up to a
critical underpressure, expanding the bubble. For $N=3000$ this critical
underpressure is indicated with an arrow, as well as the maximum radius of a
3000 electron bubble.

FIG. 2. The frequency of ripplon modes $\ell =1,..,10$ is shown as a
function of external applied pressure (not vapor pressure) for $N=1000$. The
leftmost points of the graphs start at the critical underpressure: for a
pressure more negative than this value, the bubble is unstable against
isotropic expansion. As the pressure is increased, more and more modes
obtain a vanishing frequency. On the left the deformations (exagerated) are
shown for a few of the modes.

\bigskip

\bigskip

\end{document}